%% file: arXiv-20260804.tex
\documentclass[aps,twocolumn,superscriptaddress,amsmath,amssymb]{revtex4-2}
\usepackage{graphicx}
\usepackage{subfigure}
\usepackage{multirow}
\usepackage{natbib}
\usepackage{array}
\usepackage{multirow,amssymb,amsbsy,amsmath,epstopdf}
\usepackage[dvipsnames]{xcolor}
\usepackage{ulem}
\usepackage{newtxmath}
\usepackage{float}
\usepackage{appendix}
\usepackage{titlesec}

\newcommand{\bra}[1]{\ensuremath{\left\langle #1\right|}}
\newcommand{\ket}[1]{\ensuremath{\left|#1\right\rangle}}
\usepackage[colorlinks,citecolor=blue]{hyperref}
\newcommand{\R}[1]{\textcolor{black}{#1}}
\newcommand{\B}[1]{\textcolor{black}{#1}}

\newcommand{\Tr}[1]{\mathrm{Tr}{\left[#1\right]}}

\newcommand{\out}[1]{\ket{#1}\!\bra{#1}}

\newcommand{\zl}[1]{{\color{black}{#1}}}

\begin{document}

\title{\R{Observation of quantum nonclassicality without freedom of choice \\ in a minimal causal network}}

%\iffalse

\author{Ya Xiao}
%\email{xiaoya@ouc.edu.cn}
 \affiliation{College of Physics and Optoelectronic Engineering, Ocean University of China, Qingdao 266100, China.}
 \author{Yan-Xin Rong}%
\affiliation{College of Physics and Optoelectronic Engineering, Ocean University of China, Qingdao 266100, China.}
\author{Ran He}
\affiliation{Hefei Unitary Quantum Technology Co., LTD., Hefei 230026, China}
\author{Yu~Meng}
\affiliation{Center for Macroscopic Quantum States (bigQ), Department of Physics, Technical University of Denmark, Fysikvej, 2800 Kongens Lyngby, Denmark}
\author{Yang Zhang}
\affiliation{Hefei Unitary Quantum Technology Co., LTD., Hefei 230026, China}
\author{Xiao-Ye~Xu}
\affiliation{Hefei Unitary Quantum Technology Co., LTD., Hefei 230026, China}
\author{Yong-Jian Han}
\affiliation{Hefei Unitary Quantum Technology Co., LTD., Hefei 230026, China}
\author{Zheng-Hao Liu}
\email{zheli@dtu.dk}
\affiliation{Center for Macroscopic Quantum States (bigQ), Department of Physics, Technical University of Denmark, Fysikvej, 2800 Kongens Lyngby, Denmark}
\author{Yong-Jian Gu}%
 \email{yjgu@ouc.edu.cn}
\affiliation{College of Physics and Optoelectronic Engineering, Ocean University of China, Qingdao 266100, China.}
\affiliation{Engineering Research Center of Advanced Marine Physical Instruments and Equipment (Ministry of Education), Ocean University of China, Qingdao 266100, China.}
\affiliation{ Qingdao Key Laboratory of Optics and Optoelectronics, Ocean University of China, Qingdao 266100, China}

%\fi
	
\begin{abstract}
Quantum causal networks enable tests of nonclassicality beyond Bell nonlocality while relaxing some physically unwarranted assumptions. By relaxing the freedom‑of‑choice and spacelike‑separation assumptions, the unrelated‑confounders causal networks provide a simple and robust route to certify quantum nonclassicality. Here, we implement the minimal three‑node unrelated‑confounders network in an optical experiment using two independent polarization‑entangled photon sources and an intervention at the central node, experimentally achieved with a high‑fidelity entangling measurement. We employ a causal data‑fusion protocol that combines observational and interventional data to significantly improve the protocol's noise tolerance, and observe a violation of the corresponding hybrid causal inequality by more than three standard deviations. Our results provide deeper insights into quantum nonlocality in networks and highlight the UC network as a compact, experimentally accessible platform for device‑independent quantum protocols that do not require actively chosen measurement settings.
\end{abstract}

\maketitle

%%%%%%%%%%
\iffalse
zl: The result is nice because (1) UC is much easier to realize compared to the triangle network. (2) UC's nonclassicality with relaxation is hard to calculate (1411.4648, 2404.12790), therefore, that UC is nonclassical with causal relaxation is a surprising result. (3) While a full observational scheme will require extreme visibility to warrant network nonlocality, with intervention it becomes possible. We should formulate the introduction around these points.
\fi
%%%%%%%%%%
    
\textit{Introduction.---}
\B{Bell’s theorem reveals that no local-realist model can reproduce quantum correlations ~\cite{Bell1964}, a discovery that has transformed quantum foundations into operational resources for device-independent protocols such as quantum cryptography ~\cite{ekert1991quantum,acin2007device,lu2026device}, randomness certification ~\cite{pironio2010random,liu2018device}, and communication‑complexity advantages~\cite{buhrman2010nonlocality}. }
For decades, researchers have focused on the standard Bell scenario, where observers share a single multipartite source. However, performing fully loophole-free Bell tests remains experimentally challenging, requiring strict space-like separation, near-perfect detection efficiency, and genuinely free choices of measurement settings.
Although Bell tests have closed locality and detection loopholes in photonic and superconducting systems~\cite{hensen2015loophole,giustina2015significant,shalm2015strong,big2018challenging,storz2023loophole}, \B{the freedom-of-choice assumption remains one that must be imposed rather than physically guaranteed in any laboratory experiment. This inherent tension has  motivated the search for nonclassicality tests that relax the freedom-of-choice requirement while retaining operational significance.}

\B{A promising direction lies in quantum causal networks with multiple independent sources~\cite{tavakoli2022bell}. These networks  generate correlations with no analogue in conventional Bell experiments~\cite{renou2019genuine,pozas2023proofs,renou2022network,renou2022nonlocality,boreiri2025noise,abiuso2022single}, and several experimental demonstrations have been reported~\cite{suprano2022experimental,polino2023experimental,wang2024experimental,meskine2025experimental}.
Remarkably, in some triangular causal networks, nonclassical correlations can be certified using fixed measurement settings, thereby replacing the freedom-of-choice assumption with the physically distinct condition of source independence \,\cite{renou2019genuine,pozas2023proofs}. The relaxation of measurement-choice freedom is not merely a conceptual curiosity: it is relevant for scenarios where measurement settings are not under active experimental control but are passively determined by the environment, such as in certain satellite-based quantum communication scenarios or tests of nonclassicality in naturally occurring quantum systems.}

Given its general appeal, a natural question arises: What is the simplest quantum network setting that enables nonclassicality without freedom of choice? \B{Quite surprisingly, Lauand \textit{et al.} recently showed that the answer is not the triangular network but the unrelated-confounders (UC) scenario \cite{lauand2025minimal}. The UC network requires only three observable nodes, binary outcomes, and two independent sources—making it the minimal causal structure for input-free nonclassicality tests. In comparison, the triangular network requires three sources and four-outcome measurements \cite{renou2019genuine}. 
}
 
\B{Despite the conceptual advantage, the direct observation of nonclassicality in the UC network is highly demanding \cite{lauand2023witnessing}. While a systematic framework for certifying nonclassicality under relaxed causal assumptions has been developed using Bayesian networks and linear programming~\cite{Chaves2015Unifying}, the nonclassicality in the UC network verified using only observational data \R{requires a fidelity of maximally entangled states exceeding $99.2\%$}, putting it beyond the reach of most existing experimental platforms. Lauand et al. proposed that this fragility can be overcome by introducing an intervention on the central node -- a causal operation that erases its incoming causal influences and removes the need for space-like separation~\cite{lauand2025minimal,chaves2018quantum,agresti2022experimental}.  Remarkably, the intervention-assisted protocol relaxes the \R{required fidelity to $90.8\%$}, making nonclassicality in the UC causal network feasible with current technology. 
}

\B{Here, we report the first experimental demonstration of quantum nonclassicality in the minimal UC causal network with an intervention at the center node. We generate two independent polarization-entangled photon-pair sources with the lowest fidelity of  $98.6\% \pm 0.1\%$
 and implement joint entangling projective measurement  (JEPM) 
 using tunable unitary transformations and optical Bell-state measurements. By applying a causal data-fusion protocol that combines observational and interventional data, we observe a violation of the corresponding causal inequality by over three standard deviations. Our results establish the intervention-assisted UC network as a compact, accessible platform for device-independent quantum protocols without active measurement choice, and demonstrate that nonclassicality without freedom of choice can be certified in substantially less technically demanding settings.}

\begin{figure} [htbp]
\includegraphics[width=\linewidth]{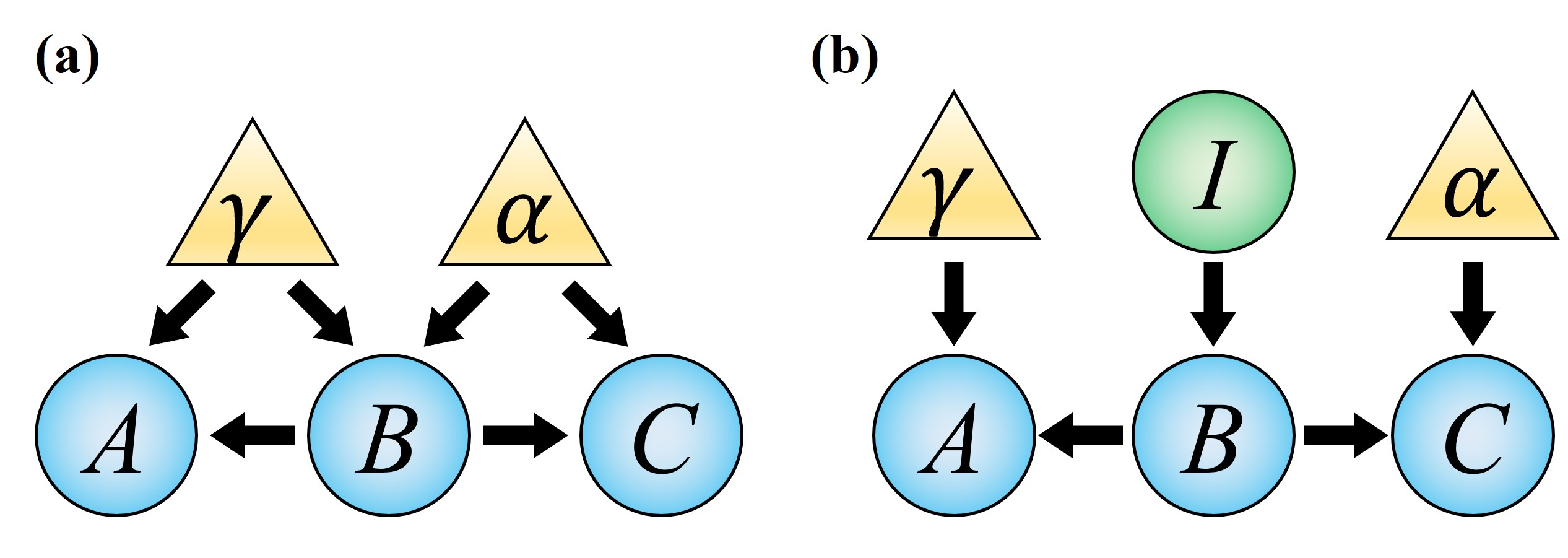}
\caption{Directed acyclic graphs of the UC scenario. (a) In the standard UC scenario, latent factors $\gamma$ and $\alpha$ act as unobserved common causes influencing the variables $A$, $B$, and $C$ (with binary outcomes $a, b, c \!\in\! \{+1, -1\}$), whose causal relations are to be inferred. In the quantum case, $A$, $B$, and $C$ are outcomes of measurements and $\gamma$ and $\alpha$ correspond to the shared quantum states $\rho_{\gamma}$ and $\rho_{\alpha}$. The value of $B$ determines the measurement settings for obtaining $A$ and $C$. (b) The UC scenario with intervention. An intervention on the central node sets the value of $B$ and removes all its other causal dependencies. Circular nodes denote observable variables; triangular nodes represent latent variables.}
\label{DAG}
\vspace{-1em}
\end{figure}

\textit{UC network.---} The UC scenario can be described by the directed acyclic graph (DAG) in FIG.~\ref{DAG}(a). The three observable nodes labeled as $A$, $B$, and $C$ represent Alice's, Bob's, and Charlie's random variables. The directed edges encode their causal relations, which are mediated by two independent latent sources $\gamma$ and $\alpha$. Binary measurement outcomes are denoted by $a$, $b$, and $c$ (with $a, b, c \in \{+1, -1\}$), each associated with its respective random variable.  Within a classical causal model, the observed distribution can be decomposed as \cite{costa2016quantum}
\begin{equation}
	P(a,b,c) = \sum_{\alpha,\gamma} P(\alpha)P(\gamma)P(a|b,\gamma)P(c|b,\alpha)P(b|\alpha,\gamma),
\end{equation}
with the independence condition $P(\alpha,\gamma) \!=\! P(\alpha) \! P(\gamma)$. In the quantum realization, $\gamma$ and  $\alpha$ correspond to shared quantum states  $\rho_{\gamma}$ and $\rho_{\alpha}$. The observed probabilities that allow for a quantum description in this scenario take the form
\begin{equation}
	P_Q(a,b,c) = \Tr{\rho_{ABC} (E_{a|b} \otimes E_b \otimes E_{c|b})}.
\end{equation}
where $ \rho_{ABC}= \rho_{\gamma}\otimes \rho_{\alpha}$.  $\{E_{a|b}\}$, $\{E_{b}\}$, and $\{E_{c|b}\}$ denote the binary positive operator-valued measures for Alice, Bob, and Charlie, respectively. Note that the model resembles that in the standard entanglement swapping experiment, but the communication from Bob to Alice and Charlie can be done with a different strategy. This choice of communication strategy is crucial for our observation of nonclassicality.

Throughout this paper, we focus on the following strategy, where the sources $\rho_{\gamma}$ and $\rho_{\alpha}$ prepare maximally entangled states $(\ket{01}+\ket{10})/\sqrt{2}$, and Bob performs a JEPM $\{E_{b}\}$ defined by $E_{+1} = \out{\psi^+_\theta}, \quad E_{-1} = \mathbb{I} - E_{+1}, $ with $\ket{\psi^+_\theta} = \sin \theta \ket{01} + \cos \theta \ket{10}$, where $\theta \in \left[ 0, \pi/4 \right]$.
He then communicates his outcome $b$ to Alice and Charlie, who set their local measurement bases accordingly. For $b = +1 (-1)$, they will measure along the $\sigma_x (\sigma_z)$ basis. 
Lauand \textit{et al.} found that \R{the strategy allows a violation of a causal inequality}, ${\Delta_{\cal N}} := {\cal N}-{\cal C_N}\leq 0$\,\cite{lauand2025minimal}, where ${\cal C_N}=3/\sqrt{2}+\sqrt{7}/6 \approx 
2.56228$ is the classical bound. \R{The definition of $\cal N$ %shall be given in Supplementary Material\,\cite{SM}. 
is given in Eq. (9) of Ref.~\cite{lauand2025minimal}.} However, the nonclassicality is fragile and prone to noise. \R{For example, a critical fidelity of~99.2\% is required for the entangled sources to manifest any quantum violation of the inequality}.

\textit{Intervention.---}To improve the noise robustness of nonclassicality in the UC scenario and render it experimentally accessible with current technology, interventions on the central node can be implemented \cite{pearl2009causality,balke1997bounds}. In the interventional case whose DAG is shown in FIG.~\ref{DAG}(b), an intervention on Bob, denoted by do($b$), effectively erases all sources of confounding and fixes his outcome to a particular value, thereby breaking the correlations between \R{Alice ($\gamma$) and Charlie $(\alpha)$}. The ability to alter the causal link makes the nonclassicality easier to detect.
Under this intervention, Alice and Charlie still select their measurement bases conditioned on the fixed value of $b$. The resulting statistics are described by the do-conditional probabilities
\begin{equation}
P_Q[a|\mathrm{do}(b)] = \Tr{\rho_\gamma E_{a|b}}, \;\; P_Q[c|\mathrm{do}(b)] = \Tr{\rho_\alpha E_{c|b}}.
% 	\begin{split}
% 		P_Q[a|\mathrm{do}(b)] &= \Tr{\rho_\gamma (E_{a|b} \otimes \mathbb{I}) }, \\
% 		P_Q[c|\mathrm{do}(b)] &= \Tr{\rho_\alpha (E_{c|b} \otimes \mathbb{I}) },
% 	\end{split}
\label{eqn:do-probabilities}
\end{equation}
whose classical counterparts are $ P[a|\mathrm{do}(b)]\!=\!\sum_{\gamma}P(\gamma)P(a|b,\gamma)$ and $ P[c|\mathrm{do}(b)]\!=\!\sum_{\alpha}P(\alpha)P(c|b,\alpha)$. 
% The corresponding correlators for Alice and Charlie are then redefined as $\expval{A}_{\mathrm{do}(b)} = \sum_a a P[a|\mathrm{do}(b)]$, and $	\expval{C}_{\mathrm{do}(b)} = \sum_c c P[c|\mathrm{do}(b)]$.

By incorporating the do-conditional probabilities, nonclassical behavior in the UC scenario involving interventions can be certified by violating the following hybrid observational–interventional nonlinear inequality\,\cite{lauand2025minimal}
\begin{align}
    \mathcal{I}(\theta) &= 2\sum_{a=c}\sqrt{P(a,+1,c)} + 4\sqrt{P(-1,-1,+1)} - G(P) \le \mathcal{C}_{\mathcal{I}},
    \label{inequality}
\end{align}
where $\mathcal{C_I} = 9/7 + 7\sqrt{6}/10 \approx 3.0004$ is the classical bound. \R{The penalty term, defined as
$G(P) := \displaystyle \sum_{a=c}\left|P(a,-1, c)-\frac{1}{4}\right|+\left|\langle A C\rangle_{-1}-\frac{1}{4}\right| + \left|\sum_{a \neq c} P(a,-1, c)-\frac{1}{4}\right|
\textstyle+\left|\langle A\rangle_{-1}+\langle C\rangle_{-1}\right| +\left|\langle A\rangle_{+1}\right|+\left|\langle C\rangle_{+1}\right|+18\left|P(b=1)-\frac{1}{4}\right|+\sum_b\left(\left|\langle A\rangle_{\operatorname{do}(b)}\right|+\left|\langle C\rangle_{\operatorname{do}(b)}\right|\right)$, incorporates the effect of marginal distributions of observables $A$, $B$ and $C$ into the classical bound.} The maximum of $\mathcal{I}(\theta)$ is found via quadratic programming to be attained at $\theta_{\mathrm{opt}} = \arctan(1/3)$ with  $\mathcal{I}(\theta_{\mathrm{opt}}) = \sqrt{10} \approx 3.1623$. Notably, the quantum violation of the hybrid inequality is robust to noise: \R{the critical fidelity of the entangled sources is relaxed to 90.8\%, making the inequality suitable for detecting nonclassicality in realistic quantum networks.}

\begin{figure*} [htbp]
	\includegraphics[width=1.02\linewidth]{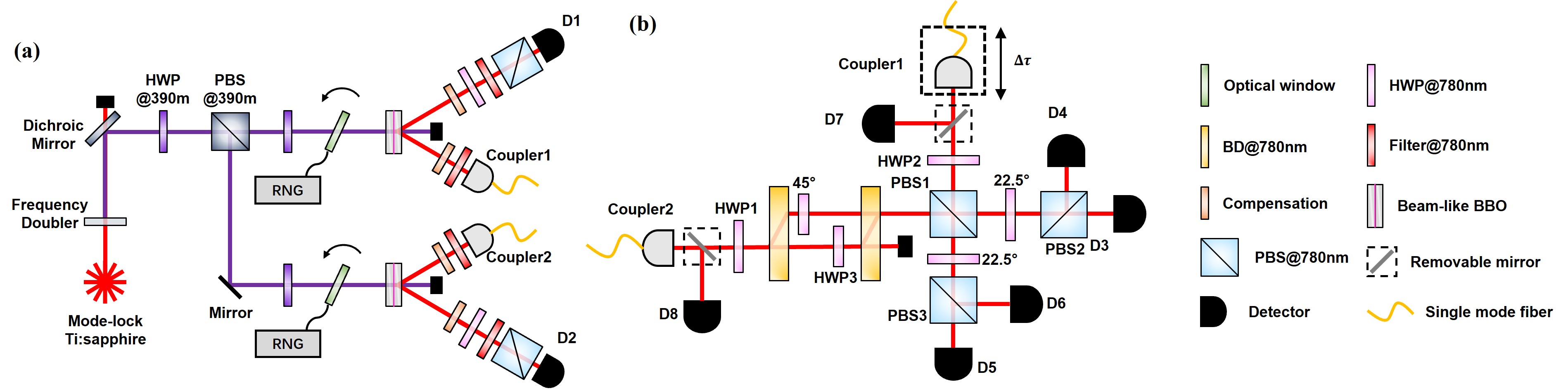}
	\caption{Experimental setup. (a) Entangled photon-pair generation. Two beam-like type-II BBO crystals, pumped in parallel by 390 nm femtosecond pulses obtained from frequency-doubled 780 nm pulses, generate two photon pairs. Each pair is prepared in the polarization-entangled Bell state $(\ket{HV}+\ket{VH})/\sqrt{2}$ after spatial and temporal compensation. Independently rotating glass plates randomize the phases of the pump pulses to suppress unwanted temporal correlations and ensure statistical independence between the two sources~\cite{wang2023certification}. (b) Bob's measurement device. The two photons sent to Bob are combined for a joint measurement, with one photon passing through a Mach-Zehnder interferometer (MZI) before both enter a linear-optical Bell-state analyzer. In the  observational scenario, Bob performs the entangling projective measurement and communicates his outcome $b$ to Alice and Charlie. They then implement single-qubit projective measurements using half-wave plates (HWPs) and polarizing beam splitters (PBSs), choosing the  $\sigma_x$ ($\sigma_z$)  basis if $b=+1$ ($b=-1$). In the interventional scenario, Bob's photons are redirected to auxiliary detectors without a joint measurement; instead, $b$ is assigned by a random number generator, while Alice and Charlie perform the same projective measurements as in the observational case.}
	\label{setup}
\end{figure*}    

\textit{Experimental setup.---}  To test quantum violations of the causal bounds in Eq.~(\ref{inequality}), we employ a photonic experimental apparatus implementing the UC scenario shown in FIG.~\ref{DAG}, enabling the generation of both observational and interventional data. Qubits are encoded in the polarization degree of freedom of photons, with $\ket{0}$ and $\ket{1}$ represented by the horizontal and vertical polarization states $\ket{H}$ and $\ket{V}$.  The entanglement-source preparation setup is shown in FIG.~\ref{setup}(a). We generate ultraviolet pump pulses at 390 nm with a pulse width of 140 fs and a repetition rate of 80 MHz by frequency doubling the output of a mode‑locked Ti:sapphire laser centered at 780 nm. The pump beam is directed to an ultraviolet polarizing beam splitter (PBS), which transmits horizontally polarized pulses  $\ket{H}$  into the upper path and reflects vertically polarized pulses $\ket{V}$ into the lower path. We use a half-wave plate (HWP) placed before the PBS to adjust the pump power in the two paths and maximize the fourfold coincidence rate. In each path, the pump pulse is rotated to a diagonally linear polarization state and then focused into a sandwich-like structure composed of two beam-like type-II $\beta$-barium borate (BBO) crystals with a true-zero-order HWP inserted between them, thereby generating polarization-entangled photon pairs in the Bell state $(\ket{HV}+\ket{VH})/\sqrt{2}$ via spontaneous parametric down-conversion \cite{pan2001experimental}. The state generated in the upper path is labeled $ \rho_{\gamma} $, while the state generated in the lower path is labeled $ \rho_{\alpha} $.

To ensure the independence of the photon sources, we insert independently driven, randomly rotating glass plates into each pump path before the crystals. The glass plates imprint random phase and amplitude modulations onto the pump pulses and effectively remove the pump correlation\,\cite{wang2023certification}. To improve the quality of the entangled photon pairs,  $\text{LiNbO}_{3} $ crystals, $\text{YVO}_{4} $ crystals, narrowband filters, and fine‑alignment translation stages are employed to suppress the spatial, spectral, and temporal distinguishability of the down‑converted photons.

\R{One photon from the entangled photon pair $\rho_\gamma$ ($\rho_\alpha$) }is sent to Alice (Charlie), who performs either $\sigma_x$ or $\sigma_z$ measurement using a combination of a HWP and a PBS. The other photon is sent to Bob, who implements the JEPM using the setup shown in FIG.~\ref{setup}(b). This setup combines two local unitary transformations, a Mach--Zehnder interferometer, and a standard Bell-state analyzer. To illustrate how the setup works, observe that the projector $E_{-1}$ can be decomposed as $E_{-1} \!=\! \out{\psi^-_\theta} \! + \! \out{\phi^+_\theta} \! + \! \out{\phi^-_\theta}$, where $\ket{\psi^-_\theta} \!=\! \cos\!\theta \ket{HV} \!-\! \sin\!\theta\!\ket{VH}$, $\ket{\phi^+_\theta} \!=\! \sin\!\theta\! \ket{HH} \!+\! \cos\!\theta \! \ket{VV}$ and $\ket{\phi^-_\theta} \!=\! \cos\!\theta \! \ket{HH} \!-\! \sin\!\theta \! \ket{VV}$. % Bob's measurement was therefore equivalent to 
Bob therefore needs to realize a projective measurement in the tilted Bell basis $\lbrace \ket{\psi^\pm_\theta},\ket{\phi^\pm_\theta} \rbrace$. He assigns outcome $b\!=\!+1$ when a coincidence event corresponding to $\ket{\psi^+_\theta}$ is detected. Other coincidence events $\ket{\psi^-_\theta}$, $\ket{\phi^+_\theta}$ and $\ket{\phi^-_\theta}$ are assigned $b\!=\!-1$\,\cite{del2024iso}.

We experimentally realize the projection on the \R{tilted Bell basis $\ket{\psi^{-}_\theta}$} using a Mach-Zehnder interferometer in the path of Coupler~2 followed by a standard Bell-state analyzer. The interferometer is fully transmissive for horizontally polarized $\ket{H}$ photons but transmits vertically polarized $\ket{V}$ photons with a reduced transmittance of $|\tan\theta|^2$. By attenuating part of the vertical component, the input state $\ket{\psi^{-}_\theta}$ is transformed to the Bell state $(\ket{HV}-\ket{VH})/\sqrt{2}$. Finally, the Bell-state analyzer consisting of a PBS, two HWPs set to $\pi/8$ at each output, and two more PBSs, implements a probabilistic projection onto $\ket{\psi^{-}_\theta}$. In this configuration, a successful projection is heralded by two‑photon coincidences between detector-pairs \R{D3 and D6 or D4 and D5} \, \cite{pittman2001probabilistic}. For projection onto the other tilted Bell basis, $\ket{\psi^+_\theta}$, $\ket{\phi^+_\theta}$, or $\ket{\phi^-_\theta} $, we apply local unitary transformations with HWP1, HWP2 and HWP3 to convert them into $\ket{\psi^{-}_\theta}$, with the specific angles used to implement the corresponding JEPM‑basis projector provided in Section A of the Supplemental Material (SM).

\zl{We apply a data-fusion protocol that combines the data from the observational runs and that from the interventional runs to calculate all the probabilities required in the inequality (\ref{inequality})}. In the observational runs, Bob performs the JEPM described above and communicates his outcome $b$ to Alice and Charlie. Based on the value of $b$, Alice and Charlie measure in the $\sigma_x$ basis for $b=+1$ and in the $\sigma_z$ basis for $b=-1$, yielding binary outcomes $a$ and $c$. The joint probability $P(a,b,c)$ is determined from fourfold coincidence between detectors D1 (D1$^{'}$), D2 (D2$^{'}$) (for Alice and Charlie), and either D3 and D5 or D4 and D6 (for Bob's projected photons). 

In the interventional runs, Bob does not perform a joint measurement but instead redirects his photons toward detectors D7 and D8. His outcome $b$ is determined by a random number generator, emulating an external intervention. Alice and Charlie continue to perform $\sigma_x$ or $\sigma_z$ measurements conditioned on the randomly assigned $b$. The do-conditional probability $P\left[a|\mathrm{do}(b)\right]$ and $P\left[c|\mathrm{do}(b)\right]$ are obtained from fourfold coincidence between detectors D1 (D1$^{'}$), D2 (D2$^{'}$), D7, and D8.

\textit{Experimental results.---} To demonstrate the quantum nonclassicality in the UC scenario, high-fidelity maximally entangled photon pairs and perfect measurements are required. The density matrices of the two entangled photon pairs $\rho_\alpha$ and $\rho_\gamma$ are reconstructed via quantum state tomography. Their fidelities with respect to the target Bell state  $(\ket{HV}+\ket{VH})/\sqrt{2}$ are  $F(\rho_\gamma) = \R{99.1\% \pm 0.1\%}$ and  $F(\rho_\alpha) = \R{98.6\% \pm 0.1\%} $, respectively, satisfying the visibility requirement derived in Ref.~\cite{lauand2025minimal}. Further details can be seen in Section B of the SM.

\begin{figure}[H]
	\includegraphics[width=\linewidth]{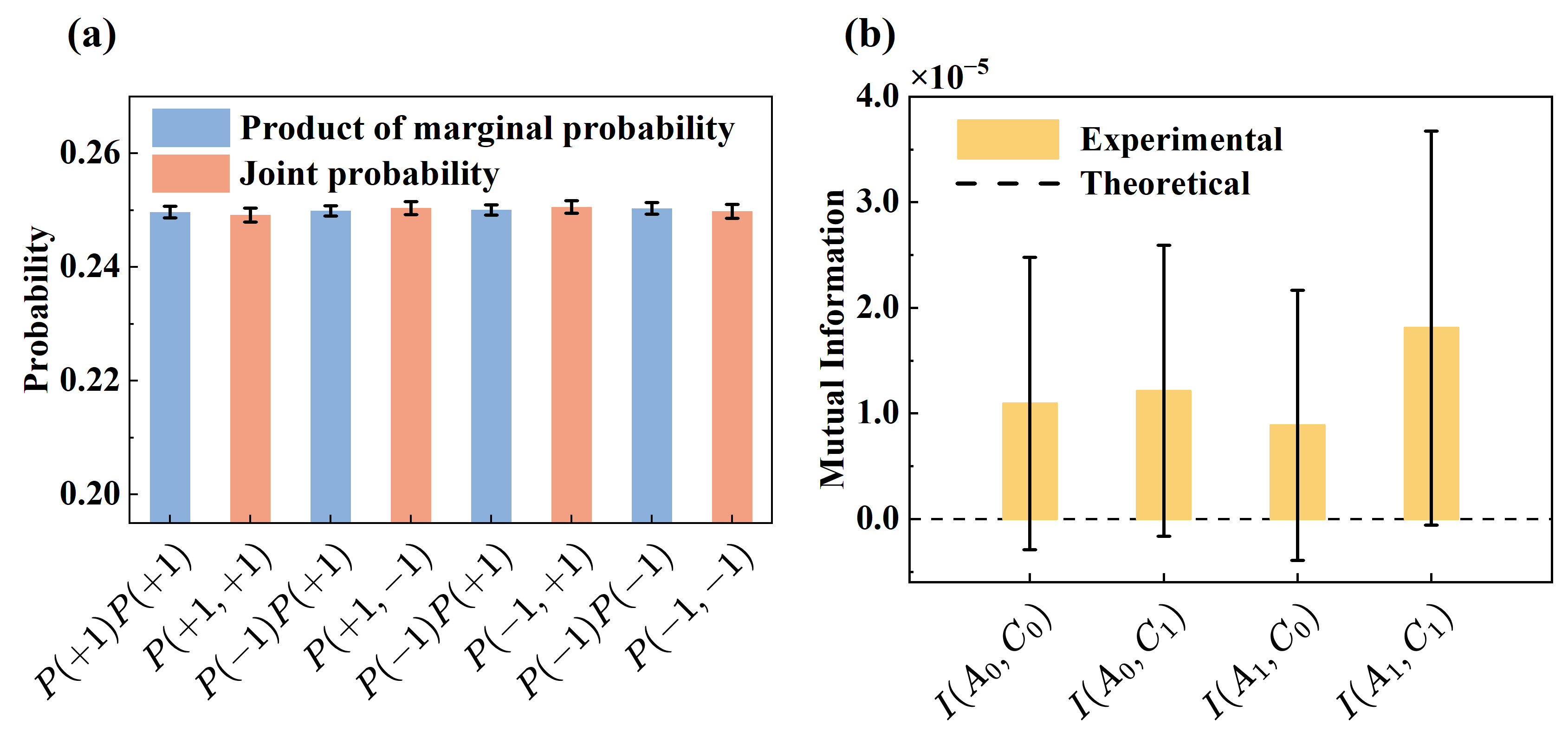}
	\vspace{-2em}
	\caption{Experimental results for source independence. (a) Test of the independence condition $P(a,c) = P(a)P(c)$ for Alice and Charlie. The measured joint probabilities $P(a,c)$ are compared with the product of the marginal probabilities $P(a)P(c)$, showing good agreement within experimental uncertainty. (b) The mutual information between Alice and Charlie. Both mean values and standard deviations are on the order of $10^{-5}$. These results provide strong evidence that the two sources can be treated as effectively independent. Error bars are estimated by assuming Poisson counting statistics.}
	\label{source_independence}
	\vspace{-1em}
\end{figure}

Source independence, a critical assumption for the UC scenario, is confirmed through multiple indicators. Firstly, the quantum state tomography of  $\rho_\gamma$ and $\rho_\alpha$ shows purities of  $0.9811\!\pm\!0.0019$ and $0.9726\!\pm\!0.0021$, respectively, both close to unity. The results reasonably rule out correlations between the sources, which would otherwise reduce the purities. Secondly, we measure the total variation distance $\displaystyle \delta\!=\!\frac{1}{2}\!\sum_{a,c}\!|p(a,c)\!-\!p(a)p(c)|$ between the joint distribution $p(a,c)$ and the product of marginals $p(a)p(c)$ and found the value to be $0.0013 \pm 0.0010$, consistent with zero (FIG.~\ref{source_independence}(a))\cite{gibbs2002choosing}. Further, the mutual information \cite{huang2022entanglement, wang2023certification, wang2025simultaneous} between Alice and Charlie, with both mean and standard deviation on the order of $10^{-5}$ (FIG.~\ref{source_independence}(b)), confirms negligible correlations between the sources. Further details are provided in Section C of the SM.

To obtain the maximal violation of Eq.~\eqref{inequality}, Bob's measurement parameter is set to $\theta = \arctan(1/3)$. We characterize Bob's JEPM using measurement tomography \cite{fiuravsek2001maximum}, which yields fidelities of  $97.6\% \!\pm\! 0.5\%$ for $\out{\psi^+_\theta}$, $96.9\% \!\pm\! 0.6\%$ for $\out{\psi^-_\theta}$, $97.2\% \!\pm\! 0.6\%$ for $\out{\phi^+_\theta}$ and $97.2\% \!\pm\! 0.5\% $ for $\out{\phi^-_\theta}$. The fidelities of our measurements well exceed the critical visibility of 87.7\% derived in~\,\cite{lauand2025minimal} using a white-noise model.  See more details in Section D of the SM.

\begin{figure} [htbp]
    \includegraphics[width=\linewidth]{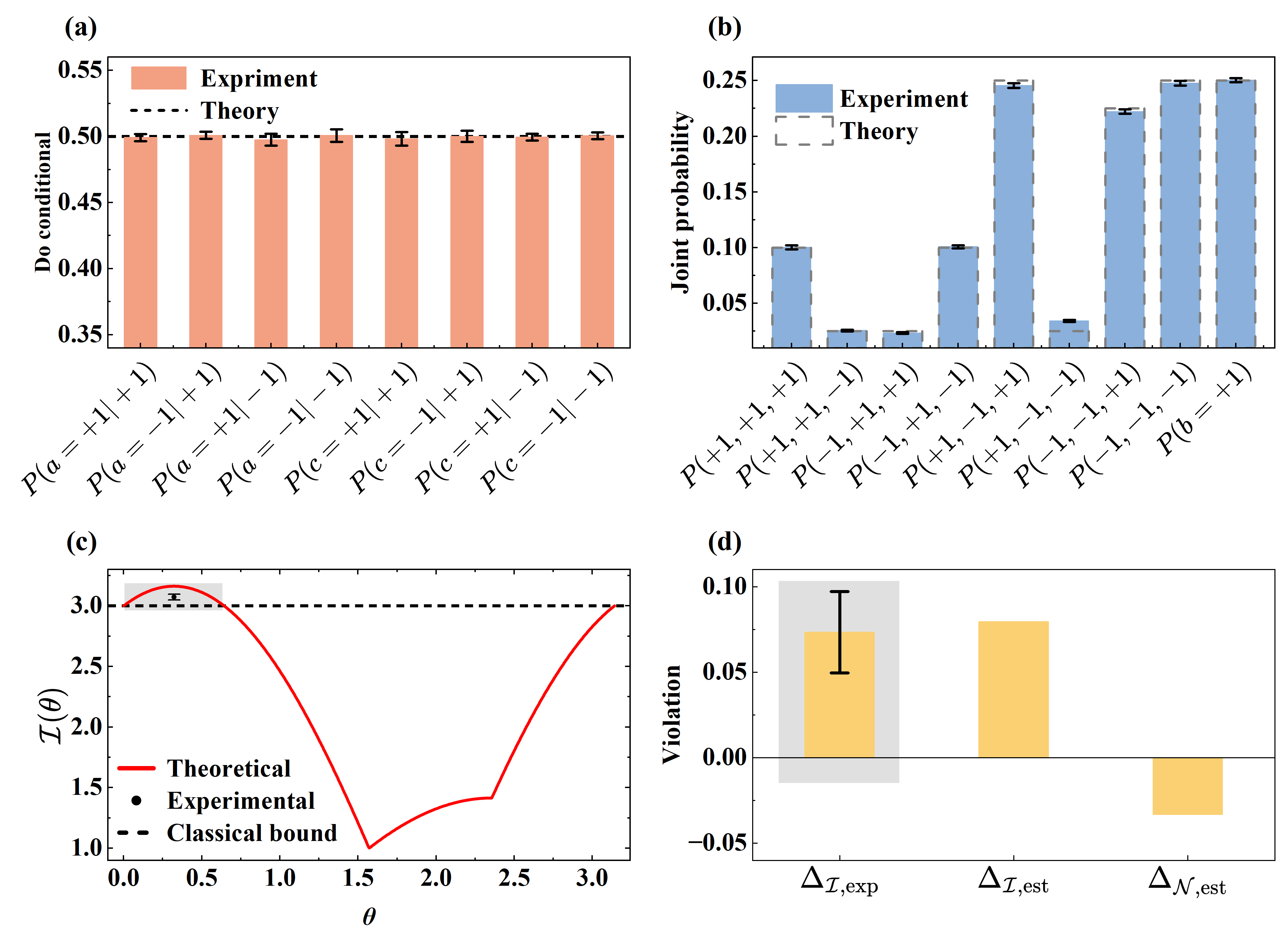}
	\caption{Measured probability distribution and violation of the inequality. (a) Joint probabilities $P(a,b,c)$ in the observational case. (b) Do conditional probabilities $P\left[a| \mathrm{do} (b)\right]$ and $P\left[c| \mathrm{do} (b)\right]$ in the interventional case. (c) Measured quantum value of Eq. \eqref{inequality} at the optimal measurement parameter $\theta_{\mathrm{opt}} = \arctan(1/3)$. The observed quantum value exceeds the classical bound by more than three standard deviations. (d) Values of $\Delta_{\mathcal{I}, \mathrm{exp}}$, $\Delta_{\mathcal{I}, \mathrm{est}}$, and $\Delta_{\mathcal{N}, \mathrm{est}}$, representing the violation of the quantum values relative to their corresponding classical bounds. The negative value of $\Delta_{\mathcal{N}, \mathrm{est}}$ indicates no violation in the non-interventional case. Error bars are estimated from Poisson counting statistics.}
    \label{p_v}
\end{figure}
We recorded the relevant four-photon coincidence counts in both the observational and interventional scenarios (FIG.~\ref{p_v}(a) and (b)) to determine the required probability distributions. In the observational case, all three parties performed projective measurements on their respective photons, yielding the joint distribution $P(a,b,c)$. In the interventional case, Bob’s outcome was replaced by an external choice of $b$, allowing for estimating the do-conditional probabilities  $P\left[a| \mathrm{do} (b)\right]$ and $P\left[c| \mathrm{do} (b)\right]$. Statistical uncertainties were estimated from Poisson counting statistics. From the measured joint probabilities, we obtain a quantum value of $\mathcal{I}(\theta)=3.0737 \pm 0.0238$, as shown in FIG.~\ref{p_v}(c). This result exceeds the classical bound of $9/7 + 7\sqrt{6}/10 \approx 3.0004$ by more than three standard deviations, clearly demonstrating nonclassical behavior in the causal network realized in our experiment.

We further compare the experimental results with those for both interventional and non-interventional inequalities, using density matrices and measurement operators reconstructed from the tomographic data in Sections B and D of the SM. As shown in FIG.~\ref{p_v}(d), for the interventional inequality (Eq. (\ref{inequality})), the experimental violation is given by  $\Delta_{\mathcal{I}, \mathrm{exp}} =\mathcal{I}_{\mathrm{exp}}-C_{\mathcal{I}}= 0.0733 \pm 0.0238$, while the tomographic estimate yields $\Delta_{\mathcal{I}, \mathrm{est}} =\mathcal{I}_{\mathrm{est}}-C_{\mathcal{I}} = 0.0794$, showing strong agreement. We also found that if intervention is not allowed, our realistic initial state cannot violate a non-interventional inequality, $\Delta_{\cal N}\le 0$, even when the measurement is assumed to be ideal. Using the density matrix from tomography (with \R{98.8\%} average fidelity to the target state), we found $\Delta_{\mathcal{N}, \mathrm{est}}\!=\!-0.0333$, \R{indicating} that no nonclassical behavior can be observed in the absence of intervention. These results provide additional support for our experimental observations and further highlight the critical role of intervention in revealing nonclassicality in the UC network.

\textit{Conclusion and Discussion.---} Our work reports an experimental observation of nonclassicality in the minimal UC network, involving only two two‑qubit entangled sources, three nodes, one binary-outcome JEPM, and two binary‑outcome qubit measurements, while \R{relaxing}  both the freedom‑of‑choice and space‑like‑separation assumptions. %The only additional assumption we employ in the experiment is the fair sampling of photon counting events. 
By further introducing an intervention on the central node,  we overcome the extreme noise sensitivity of purely observational tests, achieving a violation of  the corresponding observational‑interventional causal inequality by more than \R{three} standard deviations, certifying nonclassicality in the simplest network.

\B{In our photonic setup, the condition that the two side nodes perform measurements conditioned on the outcome of the center node is realized by data post‑processing. This is physically equivalent to an active feed‑forward strategy for the tested correlations, and can reproduce the relevant probability distributions. We verified source independence through high state purities, negligible total‑variation distance, and vanishing mutual information between the two side nodes. However, these measures do not yet constitute a physical guarantee. It remains an open question whether quantum nonclassicality in minimal networks such as the UC scenario can be certified under partially relaxed independence conditions, as recently established for bilocal networks~\cite{weilenmann2025partial}; extending such results to general causal structures would further solidify the practical foundations of network nonclassicality. }

\B{Since nonclassicality is certified here without freedom of choice, the UC network is a natural platform for device‑independent randomness certification in passive‑observation scenarios, where measurement settings are fixed by the environment, as in remote sensing or satellite‑based quantum communication. The underlying assumptions are substantially weaker than those required in instrumental or triangle scenarios \cite{sekatski2023partial,agresti2020experimental}, promising practical protocols with significantly reduced experimental demands. The intervention‑assisted methodology demonstrated here opens the door to broader explorations of quantum nonclassicality in networks where purely observational approaches remain beyond experimental reach.}

\textit{Acknowledgments---} We thank Ming-Xing Luo and Xiaotian Yang for insightful comments on the paper. We are grateful to Unitary Quantum Technology Co., Ltd. for providing the experimental platform and technical support. This work was supported by the National Key Research and Development Program of China (Grant No. 2025YFE0217700), the Shandong Provincial Natural Science Foundation (Grants No. ZR2024LLZ003, No. ZR2026LLZ016), the Fundamental Research Funds for the Central Universities (Grants No. 202364008), and the Young Talents Project at Ocean University of China (Grant No. 861901013107). \R{Z.-H.L. acknowledges support from    
    %bigQ
    Danish National Research Foundation, Center for Macroscopic Quantum States (bigQ, DNRF0142), 
    %zhenghao
    Danish e-Infrastructure Consortium (project M\AA LE, grant no.~5260-00001B),
    Villum Fonden (project FENIX, grant no.~VIL76593),
    and the MSCA European Postdoctoral Fellowships (project GTGBS, grant no.~101106833).}

\textit{Author contributions.---}Z.-H.L. and Y.X. conceived the project. Y.X. designed the experiment in discussion with Y.M., Y.-X.R. and Y.X. performed the experiment with the help of R.H., Y.Z., X.-Y.X. and Y.-J.H., Y.-X.R. and Y.X. analyzed the results with the assistance of Y.M. and Z.-H.L., Y.X., Y.-X.R. and Y.-J.G. wrote the manuscript with the input from all the other authors. Z.-H.L. and Y.-J.G. supervised the project. All authors read and approved the manuscript and discussed the experimental results.

Y. X. and Y.-X. R. contributed equally to this work.

% \bibliographystyle{apsrev4-2}	   
% \bibliography{ref.bib}
\input{arXiv-20260804.bbl}

\appendix
\newpage
\clearpage
\onecolumngrid

\renewcommand{\thefigure}{S\arabic{figure}}
\setcounter{figure}{0}
\renewcommand{\thetable}{S\arabic{table}}
\setcounter{table}{0}

\renewcommand{\thesubsection}{\Alph{section}\arabic{subsection}}

\section*{\large\centering Supplemental Material: Observation of Quantum Nonclassicality \\ without Freedom of Choice in a Minimal Causal Network}

%\section*{C. Experimental implementation and characterization of joint entangling projective measurement}
\section{Experimental implementation of joint entangling projective measurement}
\begin{figure} [h]
	\includegraphics[width=0.45\linewidth]{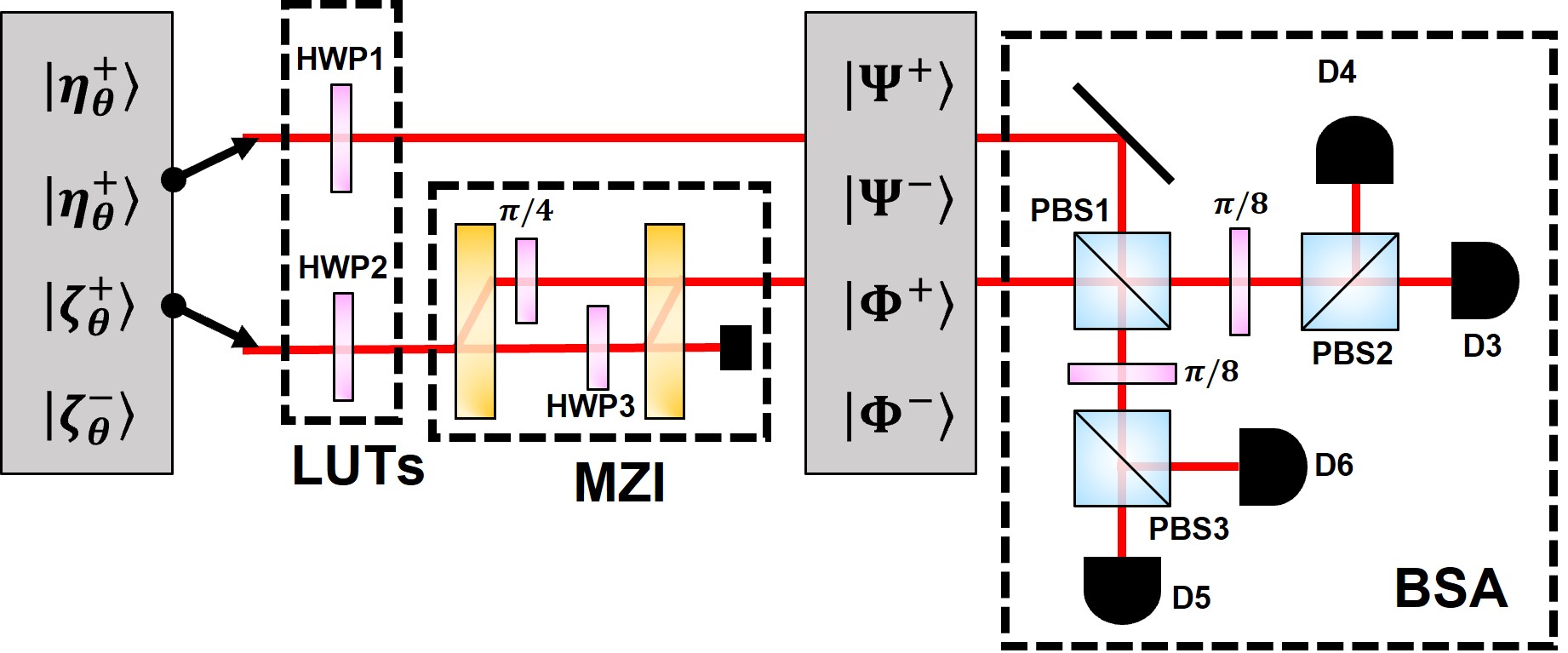}
	\caption{The JEPM is implemented using a combination of local unitary transformations (LUTs), a Mach-Zehnder interferometer (MZI), and a linear‑optics Bell‑state analyzer (BSA). The BSA can simultaneously distinguish two of the four Bell states through post‑selected coincidence events. The MZI introduces polarization‑dependent loss, enabling the realization of the partially entangled states corresponding to one eigenstate of the JEPM. The LUTs, realized using half‑wave plates (HWPs), then transform these partially entangled states into the remaining eigenstates of the JEPM.  }
	\label{JEPM}
\end{figure}
The joint entangling projective measurement (JEPM) of Bob is realized using a combination of local unitary transformations (LUTs), a Mach–Zehnder interferometer (MZI) and a linear-optics Bell-state analyzer (BSA), as shown in FIG.~\ref{JEPM}. The BSA is able to simultaneously identify two of the four Bell states $\{ \ket{\Phi^{\pm}} = (\ket{HH} \pm \ket{VV} )/\sqrt{2}, \ket{\Psi^{\pm}} = (\ket{HV} \pm \ket{VH} )/\sqrt{2} \}$ by post-selecting specific coincidence events. In our setup, coincidence events between detectors D3 and D5 (or D4 and D6) corresponds to $\ket{\Phi^+}$, whereas the response of detectors D3 and D6 (or D4 and D5) corresponds to $\ket{\Phi^-}$.

To realize the JEPM with partially entangled eigenstates, we insert a MZI before the BSA. The MZI introduces a polarization-dependent loss, completely transmitting $H$-polarized light, while partially transmitting $V$-polarized light with a transmittance of $|\tan\theta|^2$. For a complete input set consisting of the four partially entangled states $\{\ket{\eta^{\pm}_\theta} =\cos\theta\ket{HV} \pm \sin\theta\ket{VH}, \ket{\zeta^{\pm}_\theta} =\sin\theta\ket{HH} \pm \cos\theta\ket{VV}\}$, the MZI converts them into Bell states $\{\ket{\Phi^\pm}, \ket{\Psi^\pm} \}$.  After passing through the BSA, coincidence events at D3 and D5 (or D4 and D6) correspond to a projection onto $\ket{\eta^+_\theta}$, while coincidence events at D3 and D6 (or D4 and D5) correspond to a projection onto $\ket{\eta^-_\theta}$. However, among these two states, only $\ket{\eta^-_\theta}$ is an eigenstate of the JEPM, with $\ket{\eta^-_\theta}=\ket{\psi^-_\theta}$; thus, this setup implements the projector $\ket{\psi^-_\theta}\bra{\psi^-_\theta}$ with a success probability of 25$\%$. The remaining three projectors of the JEPM are obtained by applying suitable local unitary transformations using HWP1, HWP2, and HWP3. The required HWP angles for each projector are listed in TABLE.~\ref{JEPM_angles}.
\begin{table}[h]
	\centering
	\renewcommand{\arraystretch}{1.5} 
	\caption{The angles of HWP1, HWP2, and HWP3 for realizing desired JEPM projectors.}
	\label{JEPM_angles}
	\begin{tabular}{|c|c|c|c|}
		\hline
		Projector & HWP1  & HWP2  & HWP3  \\
		\hline
		$\out{\psi^+_\theta}$ & $\pi/4$ & $\pi/4$ & $\arcsin(\tan\theta)/2$ \\
		\hline
		$\out{\psi^-_\theta}$ & $0$ & $0$ & $-\arcsin(\tan\theta)/2$ \\
		\hline
		$\out{\phi^+_\theta}$ & $\pi/4$ & $0$ & $\arcsin(\tan\theta)/2$ \\
		\hline
		$\out{\phi^-_\theta}$ & $0$ & $\pi/4$ & $-\arcsin(\tan\theta)/2$ \\
		\hline
	\end{tabular}
\end{table}
\vspace{10em}

\section{More experimental results for state tomography}
\begin{figure} [h]
	\includegraphics[width=\linewidth]{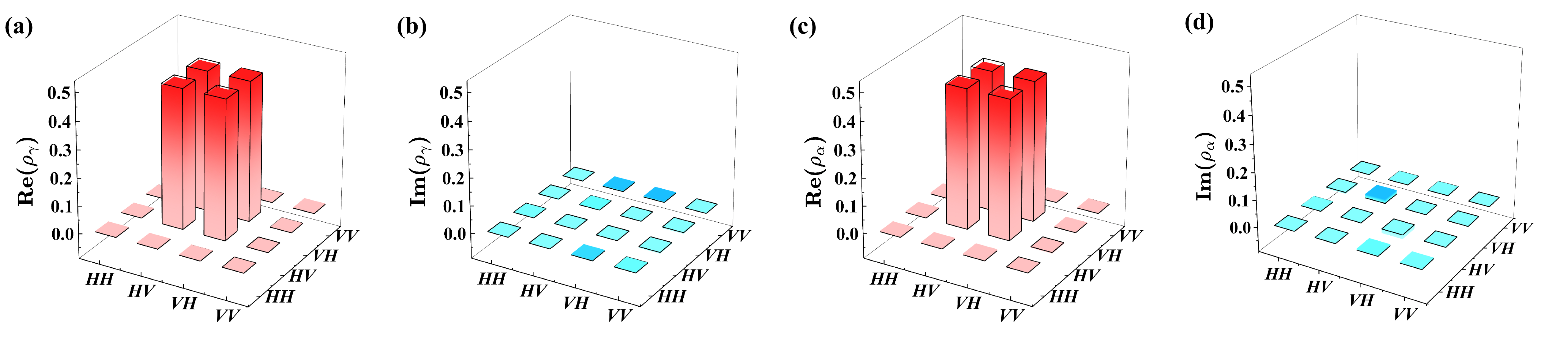}
	\caption{Tomographic results for the two polarization-entangled photon pairs. (a) and (b) plot the real and imaginary parts of $\rho_\gamma$, while (c) and (d) plot the real and imaginary parts of $\rho_\alpha$. The fidelities with respect to the Bell state $\ket{\Psi^+} = (\ket{HV}+\ket{VH})/\sqrt{2}$ are $F(\rho_\gamma) = 99.1\% \pm 0.1 \%$ and $F(\rho_\alpha) = 98.6\% \pm 0.1 \%$. Errors are estimated assuming Poisson counting statistics of the photon counts; error bars for the density matrix elements are not plotted as they are negligible (on the order of $ 10^{-3}$).}
	\label{qst}
\end{figure}

Quantum state tomography is performed on the two polarization-entangled photon pairs, labeled  $\rho_\gamma$ and $\rho_\alpha$in the main text. Their reconstructed density matrices are
\begin{equation}
\rho_\gamma = \begin{pmatrix}
					0.00283\,  & 0.00265\, +0.0007 i & 0.00226\, 	-0.00303 i & 0.00055\, -0.00123 i \\
					0.00265\, -0.0007 i & 0.49364\, & 0.49343\, 	+0.00146 i & -0.00189+0.00407 i \\
					0.00226\, +0.00303 i & 0.49343\, -0.00146 i & 	0.50067\, & 0.00162\, +0.00431 i \\
					0.00055\, +0.00123 i & -0.00189-0.00407 i & 0.00162\, -0.00431 i & 0.00286\, \\
				\end{pmatrix},
\end{equation}
\begin{equation}
	\rho_\alpha = \begin{pmatrix}
					0.00292\, & 0.00488\, +0.00197 i & 0.00599\, -0.00483 i & 0.00036\, -0.00252 i \\
					0.00488\, -0.00197 i & 0.49614\, & 0.48837\, +0.01544 i & -0.0034+0.0016 i \\
					0.00599\, +0.00483 i & 0.48837\, -0.01544 i & 0.49863\, & 0.00298\, +0.00154 i \\
					0.00036\, +0.00252 i & -0.0034-0.0016 i & 0.00298\, -0.00154 i & 0.00232\, \\
				\end{pmatrix}.
\end{equation}
The fidelities of the experimentally prepared states with respect to the target Bell state $\ket{\Psi^+}\!=\!(\ket{HV}\!+\!\ket{VH})\!/\!\sqrt{2}$ are  $F(\rho_\gamma)\!=\!99.1\%\!\pm\!0.1\%$ and $F(\rho_\alpha)\!=\!98.6\%\!\pm\!0.1\%$, respectively. The tomographic results of states $\rho_\gamma$ and  $\rho_\alpha$ are plotted in FIG.~\ref{qst}.
\vspace{-1em}

\section{More experimental results for demonstrating source independence}
We verify the independence of the two entangled photon sources through three tests: measuring the purity of the reduced states, comparing the joint probabilities with the product of the marginal probabilities, and calculating the mutual information. 

\vspace{-1.5em}
\subsection{Measuring the purities of the reduced states}
The states $\rho_\gamma$ and $\rho_\alpha$ can be considered as the reduced states of a global four-photon state. If the two entangled photon sources are correlated, the mixedness of these reduced states would increase, thereby lowering their purities. Experimentally, we obtain  $ \Tr{\rho_\gamma^2} = 0.9811 \pm 0.0019$ and $\Tr{\rho_\alpha^2} = 0.9726 \pm 0.0021$. Both values are close to 1, indicating that the states are nearly pure. This suggests that there is no significant entanglement between the two sources, supporting the assumption of their independence.
\vspace{-1.5em}

\subsection{Comparing the joint probabilities with the product of the marginal probabilities}
To directly test the independence condition, we compare the joint probabilities $p(a,c)$  with the product of the marginal probabilities $p(a)p(c)$. Specifically, $\displaystyle p(a,c)\!=\!\frac{1}{4}\!\sum_{x,z}\!p(a,c|A_x,C_z)$, $\displaystyle p(a)\!=
\!\frac{1}{2}\sum_{x}\!p(a|A_x)$ and $\displaystyle p(c)\!=\!\frac{1}{2}\sum_{z}\!p(c|C_z)$, where $p(a,c|A_x,C_z) = \Tr{(\rho_\alpha \otimes \rho_\gamma)(A_x \otimes C_z)}$  is the conditional joint probability, and the conditional marginal probabilities are given by $\displaystyle p(a|A_x)=\sum_{c,z} p(a,c|A_x,C_z)$ and $\displaystyle p(c|C_z)=\sum_{a,x} p(a,c|A_x,C_z)$. Here, $A_x$ and $C_z$ are local observables for Alice and Charlie, with $x, z \in \{0,1\}$ and outcomes \(a,c\!\in\!\{+1,-1\}\). When $x = 0$ $ (z = 0)$, Alice (Charlie) measures $A_0 = \sigma_x ~(C_0 = \sigma_x)$; when $x= 1 (z = 1)$, Alice (Charlie) measures $A_1 = \sigma_z ~(C_1 = \sigma_z)$. The deviation from the product rule is quantified by the total variation distance
\begin{equation}
 \delta = \frac{1}{2}\sum_{a,c} \left\vert p(a,c) - p(a)p(c)\right\vert.
 \end{equation}
In our experiment, we set HWP3 to $\pi/4$ and the other four wave plates (HWP1, HWP2, and the two HWPs placed after PBS1) to 0, such that Bob obtains four possible outcomes, corresponding to projections onto the eigenstates of the observable $\sigma_z \otimes \sigma_z$. Summing over Bob’s outcomes yields the conditional joint probabilities \(p(a,c|A_x,C_z)\), from which \(p(a,c)\), \(p(a)\), and \(p(c)\) are obtained according to the definitions above.The resulting value of $\delta$ is $0.0013 \pm 0.0010$, which is consistent with zero within the experimental uncertainty, thereby confirming that the independence condition is well satisfied.

\subsection{Calculating the mutual information}
As a complementary information-theoretic measure, we also calculate the mutual information (MI) between Alice and Charlie. The MI is defined as ~\cite{vedral2002role}
\begin{equation}
    I(A_x,C_z) = H(A_x) + H(C_z) - H(A_x,C_z),
\end{equation}
where $H(A_x) \!=\! -\! \sum_{a}\!p(a|A_x)\!\log{p(a|A_x)}$, $H(C_z) \!=\! -\! \sum_{c}\!p(c|C_z)\!\log{p(c|C_z)}$, and $H(A_x,\!C_z) \!=\! -\! \sum_{a,c}\!p(a,c|A_x,C_z)\!\log{p(a,c|A_x,C_z)}$ are the respective Shannon entropies.  %of the local randomness $x$, $z$ and their joint entropy, respectively. 
For perfectly independent sources, the MI vanishes. The resulting MI values are shown in FIG.~\ref{source_independence}~(b) in the main text. Both the mean values and standard deviations are on the order of $10^{-5}$ , confirming that the two sources exhibit negligible correlation and can be treated as effectively independent.

\vspace{1em}
\section{More experimental results for measurement tomography}

\begin{figure}[H]
    \centering
    \includegraphics[width=\linewidth]{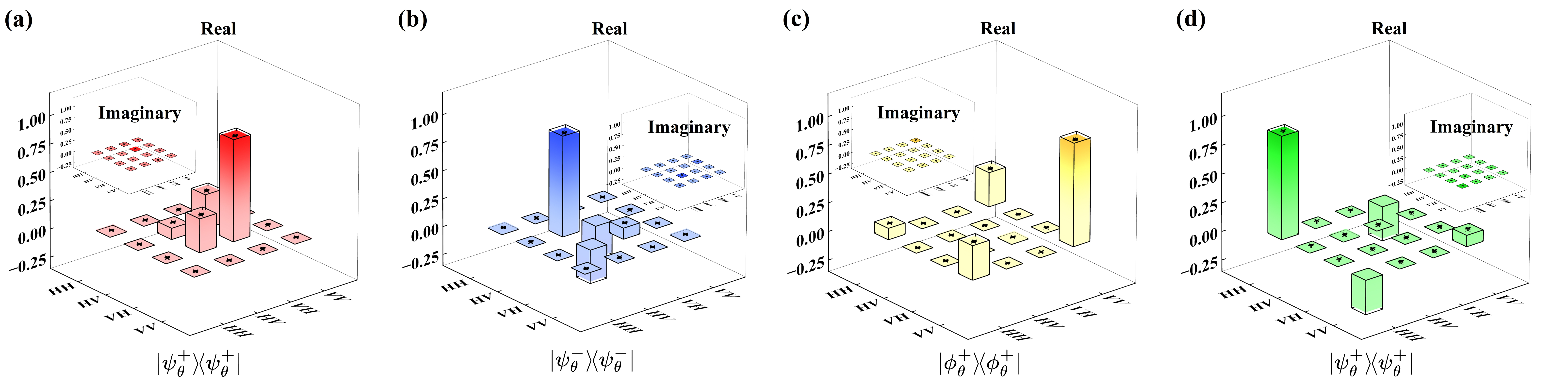}
    \caption{Tomographic results of Bob’s JEPM with $\theta = \arctan(1/3)$. Panels (a)–(d) correspond to the four projectors $\out{\psi^+\theta}$, $\out{\psi^-\theta}$, $\out{\phi^+\theta}$, and $\out{\phi^-\theta}$, respectively. The main plot in each panel shows the real part of the reconstructed operator, while the inset displays its imaginary part. Measured fidelities relative to the ideal projectors are $F(\out{\psi^+_\theta})=97.6\%\pm0.5\%$, $F(\out{\psi^-_\theta})=96.9\%\pm0.6\%$, $F(\out{\phi^+_\theta})=97.2\%\pm0.6\%$ and $F(\out{\phi^-_\theta})=97.2\%\pm0.5\%$, respectively. Errors are estimated assuming Poisson counting statistics of the photon counts.}
    \label{qmt}
\end{figure}

To characterize the JEPM implemented by Bob, we perform measurement tomography on the corresponding measurement operators. Ideally, the measurement corresponds to projections onto the tilted Bell basis $\{\ket{\psi^{\pm}_\theta}, \ket{\phi^{\pm}_\theta}\}$ with $\theta = \arctan(1/3)$, as defined in the main text. Experimentally, the measurement operators are reconstructed by preparing a total of 36 tomographically complete states as inputs and recording the corresponding four-fold coincidence. The tomographic reconstruction gives the measurement operators
\begin{equation}
	M_0 = \out{\psi^+_\theta} = \begin{pmatrix}
			0.00996 & 0.00337\, +0.00125 i & 0.00065\, +0.00349 i & -0.00108+0.00305 i \\
			0.00337\, -0.00125 i & 0.10166 & 0.28458\, +0.01496 i & -0.00049-0.00022 i \\
			0.00065\, -0.00349 i & 0.28458\, -0.01496 i & 0.88346 & 0.00304\, +0.00088 i \\
			-0.00108-0.00305 i & -0.00049+0.00022 i & 0.00304\, -0.00088 i & 0.00876 \\
		  \end{pmatrix},
\end{equation}
\begin{equation}
	M_1 = \out{\psi^-_\theta} = \begin{pmatrix}
			0.01251 & 0.00089\, -0.00092 i & 0.00107\, -0.00078 i & -0.00306-0.00221 i \\
			0.00089\, +0.00092 i & 0.87784 & -0.28173-0.00909 i & -0.00032+0.00497 i \\
			0.00107\, +0.00078 i & -0.28173+0.00909 i & 0.09632 & -0.00413-0.00274 i \\
			-0.00306+0.00221 i & -0.00032-0.00497 i & -0.00413+0.00274 i & 0.01252 \\
			\end{pmatrix},
\end{equation}
\begin{equation}
	M_2 = \out{\phi^+_\theta} = \begin{pmatrix}
			0.0938 & -0.00428-0.0011 i & -0.00054+0.00126 i & 0.28295\, +0.00827 i \\
			-0.00428+0.0011 i & 0.00907 & -0.00353-0.00208 i & -0.00524-0.00182 i \\
			-0.00054-0.00126 i & -0.00353+0.00208 i & 0.01252 & -0.00397-0.00094 i \\
			0.28295\, -0.00827 i & -0.00524+0.00182 i & -0.00397+0.00094 i & 0.88078 \\
	\end{pmatrix},
\end{equation}
\begin{equation}
	M_3 = \out{\phi^-_\theta} = \begin{pmatrix}
			0.88374 & 0.00001\, +0.00078 i & -0.00118-0.00397 i & 	-0.2788-0.00912 i \\
			0.00001\, -0.00078 i & 0.01143 & 0.00068\, -0.00379 i & 	0.00605\, -0.00292 i \\
			-0.00118+0.00397 i & 0.00068\, +0.00379 i & 0.0077 & 	0.00506\, +0.0028 i \\
			-0.2788+0.00912 i & 0.00605\, +0.00292 i & 0.00506\, 	-0.0028 i & 0.09793 \\
	\end{pmatrix}.
\end{equation}
Here, the fidelity of each measurement operators is simply defined as $F\left(M_{\mathrm{exp}},M_{\mathrm{ideal}}\right) = \Tr{\sqrt{ \sqrt{M_\mathrm{exp}} M_\mathrm{ideal} \sqrt{M_\mathrm{exp}}}}^2$, where $M_\mathrm{exp}$ and $M_\mathrm{ideal}$ represent the experimental and ideal measurement operators, respectively. The fidelities between the reconstructed operators and their ideal counterparts are $97.6\% \!\pm\! 0.5\%$ for $\out{\psi^+_\theta}$, $96.9\% \!\pm\! 0.6\%$ for $\out{\psi^-_\theta}$, $97.2\% \!\pm\! 0.6\%$ for $\out{\phi^+_\theta}$ and $97.2\% \!\pm\! 0.5\% $ for $\out{\phi^-_\theta}$. The tomographic results of Bob’s JEPM are plotted in FIG.~\ref{qmt}. 

\end{document}